\documentstyle[11pt,a4]{article}

\begin{document}

\title{Colliding plane waves with $W=M=0$}
\author{B.V.Ivanov \\
Institute for Nuclear Research and Nuclear Energy,\\
Tzarigradsko Shausse 72, Sofia 1784, Bulgaria}
\date{22 April 1997}
\maketitle

\begin{abstract}
It is shown that there are three vacuum and one electrovacuum solutions of
diagonal plane waves with $M=0$ and constant Maxwell scalars. Namely, these
are the single wave, Stoyanov, Babala and Bell-Szekeres solutions. A
comparison is made with the planar solutions of Taub.

\end{abstract}

\newpage\ 

\section{Introduction}

The line element for colliding plane waves with constant aligned
polarization has the following form in the interaction region \cite{one},%
\cite{two}: 
\begin{equation}
ds^2=2e^{-M}dudv-e^{-U}\left( e^Vdx^2+e^{-V}dy^2\right)  \label{one}
\end{equation}
where all metric functions depend on $u$ and $v$ . It is a consequence of
the requirement that the Ricci, Weyl and Maxwell scalars $\Phi _{ij}$ , $%
\Psi _i$ , $\Phi _i$ are the same at every point of the wave surfaces. If
one requires that the metric and the Maxwell tensor $F_{\alpha \beta }$
exhibit plane symmetry, then $V=0$ \cite{three},\cite{four}. We call such
solutions V-solutions. The general vacuum V-solution reads \cite{three},\cite
{five}: 
\begin{equation}
ds^2=\left( 1+Au+Bv\right) ^{-1/2}dudv-\left( 1+Au+Bv\right) \left(
dx^2+dy^2\right)  \label{two}
\end{equation}
where $A$ , $B$ are constants. Choosing them in different ways leads to a
classification theorem for such spacetimes \cite{six}. There are three
distinct V-solutions: a trivial one, depending either on $u$ or $v$ (it
reduces to Minkowski spacetime), Kasner's cosmological solution and the
static solution of Taub. The second is also the interaction field of
colliding plane shells of null dust \cite{seven}. The third may be
interpreted as the gravitational field of a massive plane \cite{eight} or a
domain wall \cite{nine}.

The electromagnetic V-solution has been found in \cite{ten}. Like the vacuum
one it is either static or cosmological and coincides with static solutions
derived in different coordinates many years ago \cite{six},\cite{eight},\cite
{eleven}. They represent the gravitation of massive charged planes.

In the present paper we study another simplified version of (1) when $M=0$ ,
calling these metrics M-solutions. They turn out to be quite similar to the
V-solutions described above. For the vacuum case we prove a classification
theorem, according to which there are only three M-solutions: a single wave,
the Stoyanov solution with $M=0$ \cite{twelve} and the Babala solution \cite
{thirteen}. It is shown in the Einstein-Maxwell case that the only
M-solution with constant $\Phi _0$ and $\Phi _2$ is the Bell-Szekeres one 
\cite{fourteen}.

The reason for such a drastic reduction in the number of solutions is the
following. The Einstein-Maxwell equations for (1) are 
\begin{equation}
\begin{array}{llll}
2U_{uu}=U_u^2+V_u^2-2U_uM_u+4\Phi _2^2 &  &  & 2U_{vv}=U_v^2+V_v^2-2U_vM_v+4%
\Phi _0^2
\end{array}
\label{three}
\end{equation}

\begin{equation}
U_{uv}=U_uU_v  \label{four}
\end{equation}

\begin{equation}
2M_{uv}=V_uV_v-U_uU_v  \label{five}
\end{equation}

\begin{equation}
2V_{uv}=U_uV_v+U_vV_u+4\Phi _0\Phi _2  \label{six}
\end{equation}

\begin{equation}
\begin{array}{llll}
2\Phi _{2v}=U_v\Phi _2-V_u\Phi _0 &  &  & 2\Phi _{0u}=U_u\Phi _0-V_v\Phi _2
\end{array}
\label{seven}
\end{equation}

In the generic case (6,7) reduce to the Euler-Darboux or to the Ernst
equation for $V$ , $\Phi _0$ , $\Phi _2$ which have numerous solutions. $M$
is then determined from (3) and (5) is a compatibility condition. However,
when $M$ vanishes and $\Phi _0$ ,$\Phi _2$ are constant, eq(3) determines
directly $V$ in terms of $U$ , given by the easily integrable (4). Then
(5,6,7) are additional constraints on $V$ which make the system highly
overdetermined, with just a few solutions. Similar considerations hold when $%
M\neq 0$ but $V=0$ .

Eqs(3-7) simplify when (1) (with $M=0$ ) is written as

\begin{equation}
ds^2=2dudv-Q^2dx^2-P^2dy^2  \label{eight}
\end{equation}

\begin{equation}
\begin{array}{llll}
P=e^{-\frac{U+V}2} &  &  & Q=e^{\frac{V-U}2}
\end{array}
\label{nine}
\end{equation}
Then (3-7) become 
\begin{equation}
\begin{array}{llll}
PQ_{uu}+QP_{uu}=-2QP\Phi _2^2 &  &  & PQ_{vv}+QP_{vv}=-2QP\Phi _0^2
\end{array}
\label{ten}
\end{equation}

\begin{equation}
\begin{array}{llll}
PQ_{uv}-QP_{uv}=2QP\Phi _0\Phi _2 &  &  & PQ_{uv}+QP_{uv}=0
\end{array}
\label{eleven}
\end{equation}

\begin{equation}
P_vQ_u+P_uQ_v=0  \label{twelve}
\end{equation}
\begin{equation}
\begin{array}{llll}
\left( Q\Phi _0\right) _u+\left( Q\Phi _2\right) _v=0 &  &  & \left( P\Phi
_2\right) _v-\left( P\Phi _0\right) _u=0
\end{array}
\label{thirteen}
\end{equation}

\section{The vacuum case}

Then $\Phi _0=\Phi _2=0$ and (13) is trivial. Eq(11) with the help of (12)
gives 
\begin{equation}
P_{uv}=Q_{uv}=\left( QP\right) _{uv}=0  \label{fourteen}
\end{equation}
which indicates that the variables separate in $P$ , $Q$ and $e^{-U}$ : 
\begin{equation}
\begin{array}{llll}
P=f_1\left( u\right) +g_1\left( v\right) &  &  & Q=f_2\left( u\right)
+g_2\left( v\right)
\end{array}
\label{fifteen}
\end{equation}
\begin{equation}
e^{-U}=f\left( u\right) +g\left( v\right)  \label{sixteen}
\end{equation}
Eq(16) is in fact (4). Eq(15) shows that for M-solutions the same property
is shared also by the square roots of the metric components. The main
equation now becomes (12): 
\begin{equation}
g_{1v}f_{2u}=-f_{1u}g_{2v}  \label{seventeen}
\end{equation}
and in addition (10) must be satisfied. There are two alternatives:

1) Suppose some of the factors in (17) vanish. There are four possible cases.

a) $f_{1u}=f_{2u}=0$

Then $f_{1,2}$ are constants and $P\left( v\right) $ , $Q\left( v\right) $
satisfy the corresponding equation in (10). We have a single wave
approaching from the right and no interaction.

b) $g_{1v}=g_{2v}=0$

Then $P\left( u\right) $ , $Q\left( u\right) $ represent a single wave
approaching from the left.

c) $f_{1u}=g_{1v}=0$

In this case $P$ is constant and from (10) we see that $Q$ is linear: 
\begin{equation}
Q=c_1+c_2u+c_3v  \label{eighteen}
\end{equation}
Rescaling $u$ and $v$ we can set $c_i=\pm 1$ . A constant $M$ appears in
(8). This is allowed because M-solutions may include also solutions with
constant $M$ since only the derivatives of $M$ enter (3-7). Having in mind
that $\sqrt{2}u=t-z$ , $\sqrt{2}v=t+z$ the signs of $c_2$ and $c_3$ should
be kept the same in order to preserve the time-like character of $t$ . Thus
we obtain the solution 
\begin{equation}
\begin{array}{llll}
Q=1\pm u\pm v &  &  & P=1
\end{array}
\label{nineteen}
\end{equation}
This is the Stoyanov solution \cite{two},\cite{twelve} 
\begin{equation}
ds^2=2\left( 1\pm u\pm v\right) ^{\frac{a^2-1}2}dudv-\left( 1\pm u\pm
v\right) ^{1-a}dx^2-\left( 1\pm u\pm v\right) ^{1+a}dy^2  \label{twenty}
\end{equation}
with $a=-1$ and \ $M=0$ . The linear terms in $e^{-U}$ cause the appearance
of impulsive components of the matter-energy tensor at the boundary of the
interaction region. Plus signs in (20) lead to negative energy, minus signs
- to the usual curvature singularity. The line element (20) may be
transformed to the cosmological Kasner metric \cite{fifteen}.

d) $f_{2u}=g_{2v}=0$

Now $Q$ is constant while $P$ is linear. Therefore 
\begin{equation}
\begin{array}{llll}
Q=1 &  &  & P=1\pm u\pm v
\end{array}
\label{twentyone}
\end{equation}
which is (20) with $a=1$ . Again we get a Stoyanov-Kasner solution.

2) Suppose none of the four factors in (17) vanishes. Then 
\begin{equation}
\frac{g_{1v}}{g_{2v}}=-\frac{f_{1u}}{f_{2u}}=c\neq 0  \label{twentytwo}
\end{equation}
\begin{equation}
\begin{array}{llll}
f_1=-cf_2+c_4 &  &  & g_1=cg_2+c_5
\end{array}
\label{twentythree}
\end{equation}
Replacing (23) in (10) gives 
\begin{equation}
\begin{array}{llll}
\left( f_1+g_1\right) f_{2uu}=c\left( f_2+g_2\right) f_{2uu} &  &  & \left(
f_1+g_1\right) g_{2vv}=-c\left( f_2+g_2\right) g_{2vv}
\end{array}
\label{twentyfour}
\end{equation}
We shall prove that $f_{2uu}=g_{2uu}=0$ and consequently $f_{1uu}=g_{1vv}=0$
. When both $f_{2uu}$ and $g_{2uu}$ do not vanish the contradiction $P=0$
follows from (24). Suppose that $f_{2uu}=0$ , $g_{2uu}\neq 0$ . Eq(24) shows
that $g_2\left( v\right) $ is constant i.e. a contradiction. The result is
similar in the other possible case. Hence $P_{uu}=Q_{vv}=0$ and following
the argument after (18) we end with 
\begin{equation}
\begin{array}{llll}
Q=1-u+v &  &  & P=1-u-v
\end{array}
\label{twentyfive}
\end{equation}
This is the solution of Babala \cite{thirteen} which describes the collision
of a plane impulsive gravitational wave with a plane shell of null matter.
One can check from (19,21,25) that (16) holds in all cases.

Thus we have proven the theorem stated in the introduction as far as the
vacuum case is concerned.

\section{The Einstein-Maxwell case}

Now we put $\Phi _0=b$ , $\Phi _2=a$ and (13) becomes 
\begin{equation}
\begin{array}{llll}
aP_v=bP_u &  &  & aQ_v=-bQ_u
\end{array}
\label{twentysix}
\end{equation}
Similar condition is found in \cite{ten} and the same reasoning leads to the
functional dependence $P=P\left( au+bv\right) $ , $Q=Q\left( au-bv\right) $
. Equations (10-13) reduce to 
\begin{equation}
\begin{array}{llll}
Q^{\prime \prime }=-Q &  &  & P^{\prime \prime }=-P
\end{array}
\label{twentyseven}
\end{equation}
where the differentiation is with respect to the corresponding argument.
Taking into account the usual boundary conditions for colliding plane
waves we find 
\begin{equation}
\begin{array}{llll}
Q=\cos \left( au-bv\right) &  &  & P=\cos \left( au+bv\right)
\end{array}
\label{twentyeight}
\end{equation}
This is the well-known Bell-Szekeres solution \cite{fourteen} , describing
the collision of two step electromagnetic waves. This ends the complete
proof of the theorem.

\section{Discussion}

There is an obvious analogy between V-solutions and M-solutions. The single
wave spacetimes correspond to each other, the Stoyanov-Kasner solution is an
analog of the Dray-'t Hooft-Kasner solution, the Babala spacetime
corresponds to the Taub static spacetime and finally, the Bell-Szekeres
solution is an analog of the Kar-McVittie solution \cite{eleven}.

The difference is that the Taub and Kar-McVittie solutions are not
interpretable as colliding plane waves. They depend on $z$ while in the
Babala and Bell-Szekeres solutions (with $a=b$ ) one metric component
depends on $z$ the other on $t$ and these metrics are not static but
resemble a peculiar kind of standing wave. Their topological structure is
very similar as noticed in \cite{two} which probably is due to the fact that
they both fall in a very restricted class of solutions.

The non-trivial Weyl scalars for (1) are $\Psi _0$ , $\Psi _2$ , $\Psi _4$ .
For V-solutions $\Psi _0=\Psi _4=0$ because their terms are proportional to
derivatives of $V$ . Consequently, these solutions are not coupled to
gravitational waves but at most to planes of null-matter \cite{seven}. For
M-solutions 
\begin{equation}
\begin{array}{llll}
\Psi _0=-\frac 1{2QP}\left( PQ_{vv}-QP_{vv}\right) &  &  & \Psi _4=-\frac
1{2QP}\left( PQ_{uu}-QP_{uu}\right)
\end{array}
\label{twentynine}
\end{equation}

\begin{equation}
\Psi _2=\frac 12M_{uv}=0  \label{thirty}
\end{equation}
which makes them complementary to the V-solutions. However, one can check
that for all M-solutions $\Psi _0=\Psi _4=0$ in the interaction region, so
they are conformally flat. The gravitational waves when present have
impulsive nature.

Finally, it should be mentioned that other M-solutions exist when $W\neq 0$
or the energy-momentum tensor includes different fields. Such is the case of
the generalized Bell-Szekeres solution \cite{two} or the collision of
classical neutrino waves \cite{sixteen}.

{\bf Acknowledgement}

This work was supported by the Bulgarian National Fund for Scientific
Research under contract F-632.

\end{document}